# Development of Lumped Element Kinetic Inductance Detectors for NIKA

M. Roesch, A. Benoit, A. Bideaud, N. Boudou, M. Calvo, A. Cruciani, S. Doyle, H.G. Leduc, A. Monfardini, L. Swenson, S. Leclercq, P. Mauskopf and K.F. Schuster for the NIKA collaboration

*Abstract*— Lumped-element kinetic inductance detectors (LEKIDs) have recently shown considerable promise as direct-absorption mm-wavelength detectors for astronomical applications. One major research thrust within the Néel Iram Kids Array (NIKA) collaboration has been to investigate the suitability of these detectors for deployment at the 30-meter IRAM telescope located on Pico Veleta in Spain.

Compared to microwave kinetic inductance detectors (MKID), using quarter wavelength resonators, the resonant circuit of a LEKID consists of a discrete inductance and capacitance coupled to a feedline. A high and constant current density distribution in the inductive part of these resonators makes them very sensitive. Due to only one metal layer on a silicon substrate, the fabrication is relatively easy.

In order to optimize the LEKIDs for this application, we have recently probed a wide variety of individual resonator and array parameters through simulation and physical testing. This included determining the optimal feed-line coupling, pixel geometry, resonator distribution within an array (in order to minimize pixel cross-talk), and resonator frequency spacing. Based on these results, a 144-pixel Aluminum array was fabricated and tested in a dilution fridge with optical access, yielding an average optical NEP of ~2 x 10-16 W/Hz^1/2 (best pixels showed NEP = 6 x 10-17 W/Hz^1/2 under 4-8 pW loading per pixel). In October 2010 the second prototype (NIKA2) of LEKIDs has been tested at the IRAM 30 m telescope. A new LEKID geometry for 2 polarizations will be presented. Also first optical measurements of a titanium nitride array will be discussed.

## I. INTRODUCTION

IN 2003, the use of the kinetic inductance effect [1] was considered for the first time as a detector application for mm and sub-mm astronomy [2]. Photons, with energy higher than the gap energy (E=h·ν>2Δ), break cooper pairs in a superconducting film. This leads to an increase in number of quasi particles, which changes the surface reactance (kinetic inductance) of the superconductor (kinetic inductance effect).

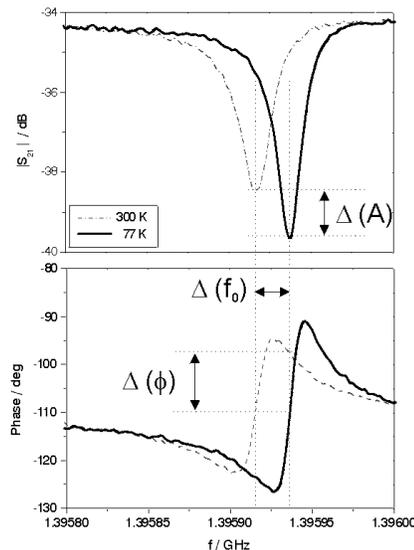

Fig. 1. Principle of KIDs. Solid lines: under dark conditions; dashed line: optical load of 300 K.

One possibility to make advantage of this effect is to use a superconducting resonant circuit as detecting element. A measurement of such a resonator coupled to a transmission line is shown in Fig. 1. An illumination of the detector leads to a shift in resonance frequency $f_0$, which can be measured in a change in amplitude and phase. One advantage of KIDs is the relative easy fabrication process. Due to only one metallization layer on a substrate, it is less complicated compared to bolometer fabrication. Another advantage of KIDs represents the readout system. Frequency multiplexing allows the readout of a large number of resonators. Packed in a limited bandwidth, a single transmission line is sufficient to read out several hundreds of pixel [3,4].

M. Roesch, S. Leclercq and K.F. Schuster are with IRAM (Institut de Radioastronomie Millimetrique), St. Martin d'Heres, France (e-mail: roesch@iram.fr, leclerqc@iram.fr, schuster@iram.fr).

A. Benoit, A. Bideaud, N. Boudou, M. Calvo and A. Monfardini are with the Institut NEEL, Grenoble, France (e-mail: alain.benoit@grenoble.cnrs.fr, aurelien.bideaud@grenoble.cnrs.fr, nicolas.boudou@grenoble.cnrs.fr, martino.calvo@grenoble.cnrs.fr, alessandro.monfardini@grenoble.cnrs.fr).

L. Swenson was with the Institut NEEL, Grenoble France. He is now with Caltech, Pasadena, CA, USA (e-mail: swenson@astro.caltech.edu).

A. Cruciani is with the Dipartimento di Fisica Universita di Roma, Rom, Italy (e-mail: angelo.cruciani@grenoble.cnrs.fr).

S. Doyle and P. Mauskopf are with the Cardiff School of Physics and Astronomy, Cardiff University, Cardiff, UK (e-mail: simon.doyle@astro.cf.ac.uk, philip.mauskopf@astro.cf.ac.uk).

H. G. Leduc is with the Jet Propulsion Laboratory, California Institute of Technology, Pasadena, CA 91109 (e-mail: Henry.G.Leduc@jpl.nasa.gov).

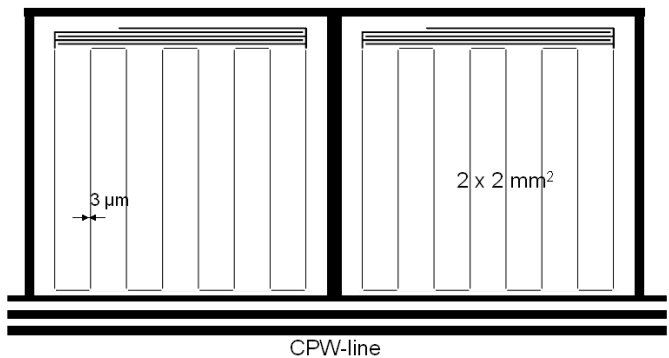

Fig. 2. Schematic of 2 LEKIDs. The resonators are coupled to a cpw-line. The line width is 3 μm and the pixel size is around 2x2 mm^2.

In Fig 2, one resonator design, a so-called lumped element kinetic inductance detector (LEKID) is shown. This type of resonator was first proposed as alternative to the quarter wavelength resonators in 2007 [5]. Compared to quarter wavelength resonators [6], this type consists of a long meandered line, the inductive part, and an interdigital capacitor. A very high and constant current density over the whole length of the meander makes this part a very sensitive direct detection area. The optical efficiency can to be optimised by changing the grid of the meander geometry. Therefore, there are no lenses or antenna structures necessary to couple the incoming microwaves into the resonator. The resonators are surrounded a ground plane to reduce the cross talk between the pixels.

The NEEL IRAM KIDs array (NIKA) is a collaboration of several groups to develop a multi pixel camera based on kinetic inductance detectors for the IRAM 30m telescope located in Spain. NIKA saw first light in October 2009 [7]. A second test run with a dual band optic took place in October 2010 [8]. Here we present the development of a LEKID array for NIKA. Measurement results of electrical and optical characterization are demonstrated in this paper.

## II. LAB MEASUREMENTS

### A. Electrical and optical characterization

We recently tested LEKID arrays with 132 pixels (see Fig. 3). The used aluminum is deposited via magnetron dc sputtering on high resistivity silicon substrate. The thickness of the aluminum is 20 nm. The arrays have been characterized electrically and optically. The arrays are measured in an H3H4-dilution cryostat with optical window. The base temperature is 80 mK. In the cryostat several filters are mounted to provide the frequency bandwidth of the 2 and 1.3 mm band. A polarizer at the 80 mK stage allows the measurement of 2 arrays at the same time, one for the 2 mm and a second for the 1.3 mm band. With this configuration the pixels respond to only one polarization of the signal, which means that the response of the detector is smaller. In Fig. 4 a VNA scan of the scattering parameter S21 of a 132-pixel aluminum array is shown. One of the problems is still the cross-talk between the pixels. There are lots of double resonances and a non-homogenous distribution which is caused mainly by the cross-talk. The variations in attenuation along the scan are due to standing waves along the feedline and the gain of the cold amplifier. The intrinsic quality factor $Q_0$ was measured to be around 100 000 under an optical load of 300 K.

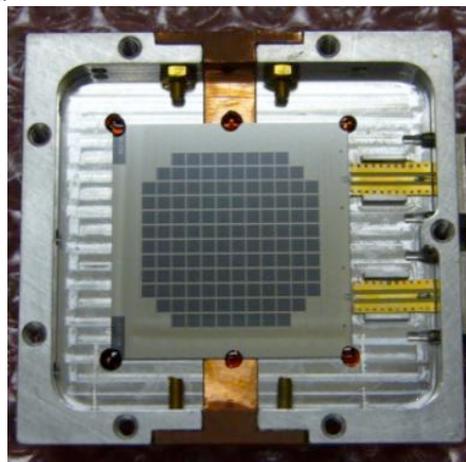

Fig. 3. Picture of a mounted 132-pixel aluminum array for the 2 mm band of NIKA.

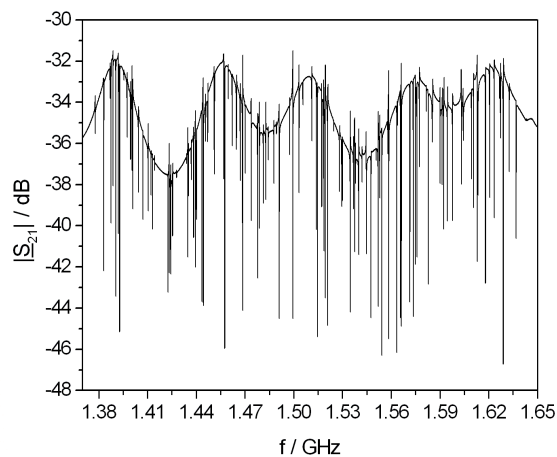

Fig. 4. VNA scan of the transmission scattering parameter S21 of the 132-pixel aluminum array at 100 mK.

To characterize the LEKIDs optically, a so-called sky simulator built at the Institute NEEL is used [9]. A picture of the simulator is shown in Fig. 5a. It is a pulse tube cryostat with a 24 cm absorber cold plate in it. The absorber can be cooled down to 50 K, which correspond almost to the background we have at the telescope. The simulator is equipped with a resistance to heat the cold plate. In Fig 5b the whole setup is shown. We use a high ε ball that is mounted on a XY-arm in front of the sky simulator. This setup allows simulating on the fly measurements as they are done at the telescope.

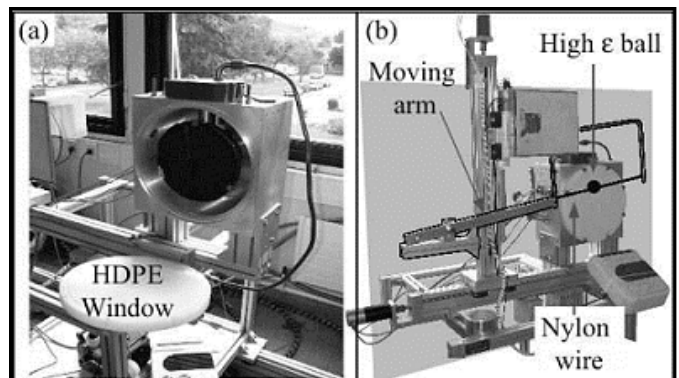

Fig. 5. Picture of the sky simulator used for the measurements in the lab. a) open pulse tube cryostat with view to the 24 cm absorber cold plate; b) complete system with high ε ball for planet simulation.

The second telescope run of NIKA took place in October 2010. More information on the telescope run can be seen in [8]. The noise spectrum shown in Fig. 6 was taken at the telescope under excellent sky conditions. The spectrum is very flat ($\sim f^{-0.15}$). The noise increases below 0.2 Hz due to sky fluctuations. The sensitivity was calculated giving an average NEP = $2.3 \cdot 10^{-16}$ W/√Hz @ 1 Hz per pix and a NEFD = 37 mJy√s per beam for the LEKIDs at 2 mm. For some best pixels a NEP of $6 \cdot 10^{-17}$ W/√Hz @ 1 Hz per pix was calculated. Remember that all the tests have been done with a polarizer that decreases the signal by a factor of 2.

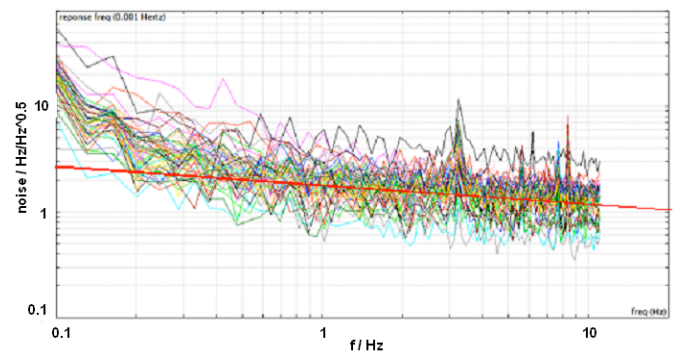

Fig. 6. Noise spectrum taken at the IRAM 30 m telescope with LEKIDs at 2mm.

### B. Optical absorption

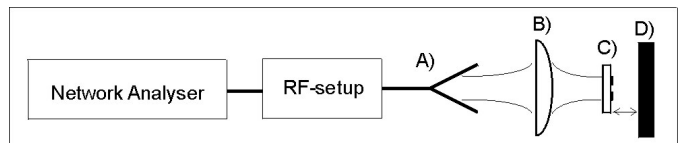

Fig. 7. Schematic of the setup for the measurement of the optical absorption of LEKIDs at room temperature. A) corrugated feed horn, B) corrugated lens, C) sample, D) back-short cavity.

To optimize the optical coupling of the LEKID structure we did room temperature measurements with the setup shown in Fig. 7. It consists of a network analyzer, an rf-setup to provide the frequency band of 120 to 180 GHz, a feed horn, a corrugated lens and the sample with its back-short. This reflection measurement setup allows tests at room temperature, which saves much time compared to cryogenic measurements. The resistivity of the aluminum was adapted to room temperature by changing the thickness by the residual resistance ration (RRR) to measure under the same conditions.

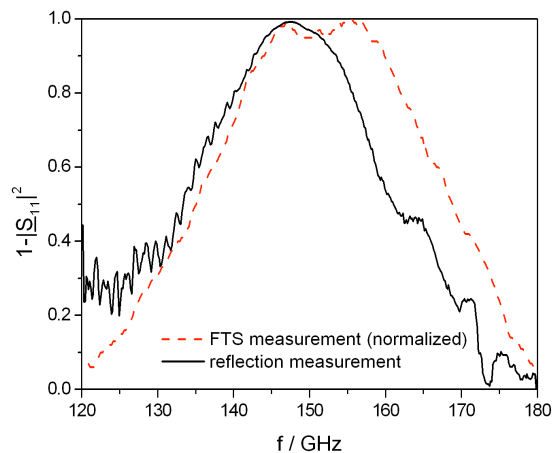

Fig. 8. Comparison of the reflections measurement results (solid black line) and FTS measurement with a Martin-Puplet interferometer (dashed red line).

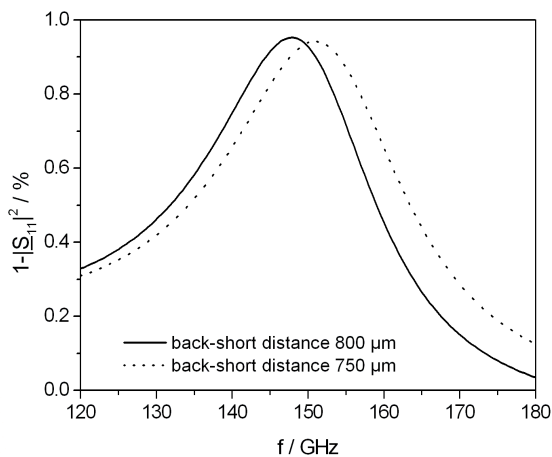

Fig. 9. Simulation of the back-short dependence of two identical structures. Solid line: back-short of 800 μm; dotted line: back-short 0f 750 μm.

Fig. 8 shows the results of the reflection measurements of a 132-pixel aluminum array (see Fig. 3). The thickness of the aluminum is 90 nm and the back-short distance is 800 μm. The absorbed frequency band is well centered at 150 GHz with a 3dB band of around 30 GHz. For comparison a second measurement, done with a Martin-Puplet interferometer at low temperatures has been done. The result is also shown in Fig. 8. The two measurements show a good agreement. The frequency band of the Martin-Puplet measurements is slightly shifted to higher frequency. The explanation for that is a smaller back-short distance of 750 μm. This smaller back-short was found by simulation and the room temperature measurements to guarantee an optimal absorption. To demonstrate the back-short dependence of the absorption, simulations have been done with a transmission line model. The result of two simulations for identical structures but different back-short is shown in Fig. 9. The results show that the characterization of the optical coupling at room temperature is a feasible alternative to the time intensive cold measurements. More information about the absorption measurements at room temperature and the transmission line model for these structures will be published elsewhere [10].

### III. LEKID-Geometry for 2 Polarizations

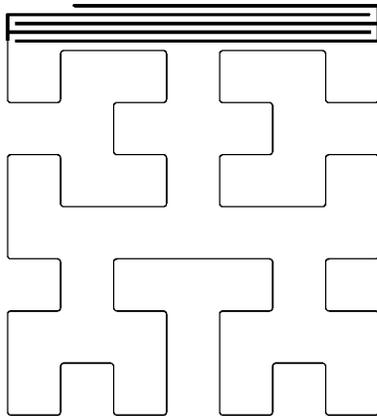

Fig. 10. The picture shows a design for a LEKID that is sensitive to 2 polarizations. The Structure is a Hilbert curve of the $3^{rd}$ order.

In the next setup of the NIKA cryostat, the polarizer is replaced by a dichroic. This allows the use of pixels that are sensitive to 2 polarizations, which increases the optical response of the detector. For this application a symmetric design with a constant filling factor over the whole direct detection area is needed to guarantee the same absorption for the 2 polarizations. One possible design is shown in Fig. 9. It is a so-called Hilbert fractal curve of the $3^{rd}$ order. This geometry is a well-known for patch antennas [11]. However for the LEKID application we use the geometry as a direct absorption area. In this case the antenna behavior is less important. It is a geometry that fulfills the conditions for the 2-polarization detection mentioned above [12].

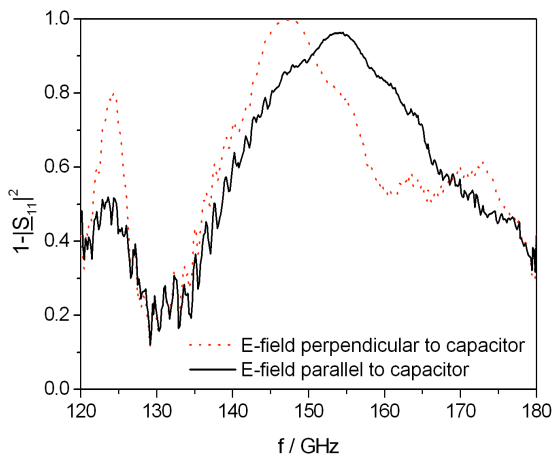

Fig. 11. The plot shows the absorption in the two polarizations of the geometry shown in Fig. 9. The measurements have been done with at room temperature with the setup shown above.

In Fig. 11 the reflection measurement results for the Hilbert geometry is shown. A back-short distance of 700 µm was found to guarantee an optimal optical absorption. The results show an almost identical absorption for the 2 polarizations. Around 150 GHz a maximum absorption of almost 100 % was measured.

After the room temperature measurements we mounted a 132-pixel array into the cryostat to characterize the detectors under real observing conditions. The results of the sensitivity of the 2-polarization sample are shown in Fig. 12. Three optical measurements have been done. The first one was done without polarizer in front of the cryostat (black squares). The second one has been done with a polarizer in front of the cryostat with a vertical orientation (blue triangles). For the third one, the polarizer was mounted with a horizontal orientation (red stars). The results show clearly almost the same sensitivity for both polarizations. Assuming that the noise is the same for both polarizations, the responses are almost identical. The better NET without the polarizer is due to the same noise level but a much higher response. The green line in Fig. 11 indicates the average value of the NET for all the pixels of the sample. It was calculated to be around 2.5 mK/Hz^0.5 @ 1 Hz per beam. Some best pixels even have shown a NET = 0.8 mK/Hz^0.5 @ 1 Hz per beam.

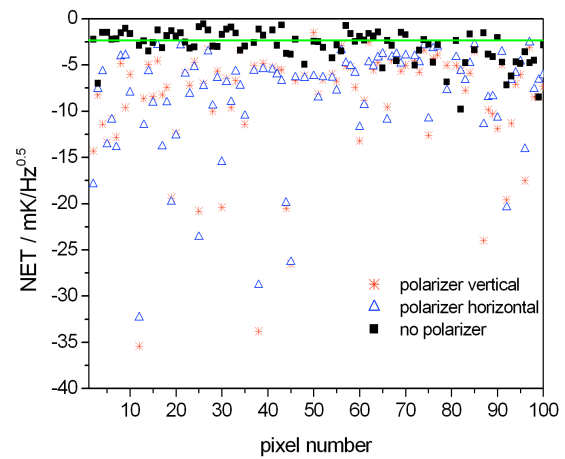

Fig. 12. Sensitivity measurement for a 132-pixel array sensitive to 2 polarizations. The measurements have been done with and without a polarizer in front of the cryostat. Black squares: no polarizer; blue triangles: polarizer horizontal; red stars: polarizer vertical.

### IV. Titanium Nitride LEKIDs

In 2010, titanium nitride was proposed as promising new material for KIDs [13]. The kinetic inductance is higher compared to aluminum, which increases the response of the detector. Moreover, the normal resistivity is higher, allowing thicker films and larger lines, which decreases problems due to lithography. We designed a mask, adapted to the much higher resistivity and kinetic inductance, for a 132-pixel TiN array for 1.3 mm. The TiN was deposited via sputtering at JPL by H. G. Leduc on a 350 µm thick silicon substrate. The film thickness is about 70 nm with a square resistance of around 30 ohms per square. The Tc was measured to be around 1.25 K. We mounted the array in the cryostat and measured it under different optical loads at 80 mK. The array was mounted at the 2mm band instead of the 1.3 mm band. The results of these measurements are shown in Fig. 13 for one resonator.

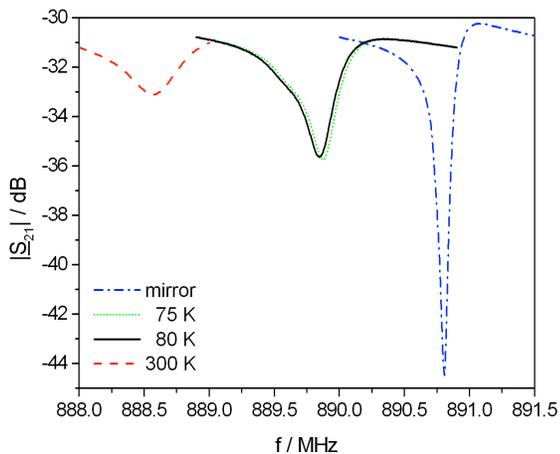

Fig. 13. Response of a TiN LEKID resonator to different optical loads. Dashed red line: 300 K; solid black line: 80 K; dotted green line: 75 K; dashed dotted blue line: mirror in front of cryostat (cryostat closed).

The shift in frequency, comparing a closed cryostat and a 300 K load, is around 2 MHz. The response of the TiN LEKIDs is around a factor of 10 higher compared to the usual aluminum arrays (200 KHz). Using the sky simulator we measured a shift in frequency of 30 kHz between 75 and 80 K background loading (see Fig. 12). The problem with this big response is that the resonances are almost destroyed under a 300 K load. Also the resonances have been found at lower frequencies than expected (700-900MHz). The cold amplifier we have used is optimized for the 1.2 – 1.8 GHz band. From the amplifier specifications, the noise is 4 to 5 times higher when at 900 MHz, and even worse below. There is thus room for improving just by using an amplifier designed for lower frequencies. In order to study the sensitivity under smaller loading, we have reduced the optical load from around 6 pW to 0.1 pW by applying a diaphragm to the pupil. A first estimation gives NEPs per pixel better than $5 \cdot 10^{-17}$ W/√Hz @ 1 Hz. This is already an encouraging result for a first test with TiN, that will surely improve considering that we are sensitive to only one polarization with the meandered geometry, and we are well out of the amplifier bandwidth.

## V. CONCLUSION

We demonstrated in this paper the high potential of lumped element kinetic inductance detectors as mm-wave detectors, in particular at 2 mm wavelength. The results of the two telescope runs show that we are competitive now to existing bolometer arrays. The good agreement between the observations at the telescope and the measurements done with the lab setup, validates the used sky simulator. The new NIKA cryostat setup with a dichroic filter allows now measuring the optical response of pixels that are sensitive to two polarizations. Some best pixels of the presented design in this paper show a sensitivity of NET = 0.8 mK/Hz^0.5 @ 1 Hz per beam. This is the best result achieved with kinetic inductance detectors, in particular for LEKIDs, at 2 mm so far. The first measurements with titanium nitride resonators show very promising results regarding the sensitivity of the pixels. Future work will address remaining issues such as pixel cross talk.

We are also investigating in aluminum and TiN arrays for 1mm. First samples have been measured but the sensitivity is not satisfactory yet.

Currently we are able to readout 112 pixels in a 230 MHz bandwidth. We are currently working on readout electronics that will allow the readout of more pixels in a larger bandwidth.

A third telescope run at the IRAM 30m is foreseen for September 2011.

ACKNOWLEDGMENT

This work was supported in part by the French National Research Agency Grant No. ANR-09-JCJC-0021-01.

Part of this work was supported by the Keck Institute for Space Studies (KISS).